\documentclass[article,twocolumn,superscriptaddress,amsmath,amssymb,prb]{revtex4-1}
\usepackage{amsmath, amssymb}
\usepackage{graphicx}
\newcommand{\ket}[1]{| #1 \rangle} 

\newcommand{\abs}[1]{|#1|}

\newcommand{\circled}[1]{\raisebox{.5pt}{\textcircled{\raisebox{-.9pt} {#1}}}}

\DeclareMathOperator{\Rre}{Re}

\begin{document}
\title{\sffamily Observation of a dissipative phase transition in a one-dimensional circuit QED lattice} 

\author{Mattias Fitzpatrick}
\affiliation{Department of Electrical Engineering, Princeton University, Princeton, NJ 08540, USA}
\author{Neereja M. Sundaresan}
\affiliation{Department of Electrical Engineering, Princeton University, Princeton, NJ 08540, USA}
\author{Andy C.\ Y.\ Li}
\affiliation{Department of Physics and Astronomy, Northwestern University, Evanston, IL 60208, USA}
\author{Jens Koch}
\affiliation{Department of Physics and Astronomy, Northwestern University, Evanston, IL 60208, USA}
\author{A.\ A.\ Houck}
\affiliation{Department of Electrical Engineering, Princeton University, Princeton, NJ 08540, USA}

\date{July 22, 2016}

\maketitle 
\textbf{\sffamily
	Condensed matter physics has been driven forward by significant experimental and theoretical progress in the study  and understanding of equilibrium phase transitions based on symmetry and topology.  However, nonequilibrium phase transitions have remained a challenge, in part due to their complexity in theoretical descriptions and the additional experimental difficulties in systematically controlling  systems out of equilibrium.   Here, we study a one-dimensional chain of $72$ microwave cavities, each coupled to a superconducting qubit, and coherently drive the system into a nonequilibrium steady state. We find experimental evidence for a dissipative phase transition in the system in which the steady state changes dramatically as the mean photon number is increased.  Near the boundary between the two observed phases, the system demonstrates bistability, with characteristic switching times as long as $60$\,ms -- far longer than any of the intrinsic rates known for the system.  This experiment demonstrates the power of circuit QED systems for studying nonequilibrium condensed matter physics and paves the way for future experiments exploring nonequilbrium physics with many-body quantum optics.}

Over the past decades, there has been remarkable progress in studying both real and synthetic quantum materials. Advances in nanoscale fabrication and cryogenics have allowed for exquisite control of electronic systems -- unlocking strongly correlated electronic states and topological materials \cite{Hasan2010}. Simultaneously, the ability to model desired Hamiltonians with ultra-cold Fermi and Bose gases has allowed unprecedented access to synthetic material properties \cite{Zwierlein2005}. As a whole, much of the development of condensed matter physics has focused on the study of (quasi-)equilibrium physics, which is more accessible both experimentally and theoretically. However, the constant presence of dissipation, noise, and decoherence belie the fact that, ultimately, the world is nonequilibrium.

A phase transition indicates a sometimes sudden change in the physical properties of a system as a function of some external system parameter. Thermal phase transitions are well-understood in the context of statistical mechanics and occur when the free energy becomes nonanalytic. At zero temperature, the role of quantum fluctuations gives rise to a new set of quantum phase transitions which involve a sudden change in the ground state of a Hamiltonian $H$; a phase transition occurs when the gap between the first excited state and the ground state closes. These concepts need to be extended to consider nonequilibrium steady states, as the system is no longer in its ground state but rather in a state that balances drive and dissipation.  In a dissipative phase transition, the steady state abruptly changes as a system parameter is varied\cite{Kessler2012}. Whenever the system is describable in terms of a Lindblad master equation\cite{Breuer2007}, $\dot{\rho}=\mathbb{L}\rho$, then such a transition is signalled by the closing of the lowest excitation gap in the  spectrum of the Liouvillian superoperator $\mathbb{L}$.

In recent years, interacting photons have  emerged as an excellent candidate for studying nonequilibrium condensed matter physics due to the lack of particle number conservation\cite{Houck2012}.  In cavity quantum electrodynamics, strong coupling between atoms and a cavity can mediate effective photon-photon interactions\cite{Birnbaum2005, Lang2011,Hoffman2011}.  Arrays of coupled microwave\cite{Underwood2012} or optical\cite{Jacqmin2014} cavities can be fabricated by conventional lithographic techniques, and the competition between on-site interactions and hopping between neighboring cavities can give rise to quantum phase transitions of light\cite{Greentree2006,Hartmann2006,Angelakis2007}. A wide range of many-body effects have been predicted in these systems, including a Mott insulator-superfluid phase transition\cite{Greentree2006,Hartmann2006,Angelakis2007} and fractional quantum Hall-like states of light\cite{Cho2008,Petrescu2012, Anderson2016}.  Experiments on small systems have demonstrated low-disorder lattices\cite{Underwood2012}, a dynamical quantum phase transition in a cavity dimer\cite{Raftery2014}, and chiral ground state currents in a cavity trimer\cite{Roushan2016}. Circuit QED lattices are inherently open systems, with dissipation an ever-present force that leads both to qubit relaxation as well as the inevitable loss of photons from microwave cavities.  While dissipation presents an obstacle for quantum information processing, it is of fundamental interest in the study of nonequilibrium phase transitions.  Just as excitations inevitably leak from the system, it is easy to add photons back and to drive into a steady state, making these systems particularly amenable to the study of dissipative phase transitions.

In this paper we present experimental evidence for a dissipative phase transition in a circuit QED lattice.  We observe that at drive frequencies between the low-power resonance frequencies of the system, there exists a region of hysteresis and bistability where the steady state of the system  switches stochastically between two states $\rho_1$ and $\rho_2$. By determining the corresponding switching rates, we can obtain the so-called asymptotic decay rate which characterizes the closing of the spectral gap of $\mathbb{L}$.  At the transition between the two states, the characteristic switching times become exceptionally long, a key characteristic of a dissipative phase transition.  A similar observation has recently been made in a single cavity system with multiple qubits\cite{Fink2016}. 

\begin{figure}[!t]
	\begin{center}
		\includegraphics[width=1.0\columnwidth]{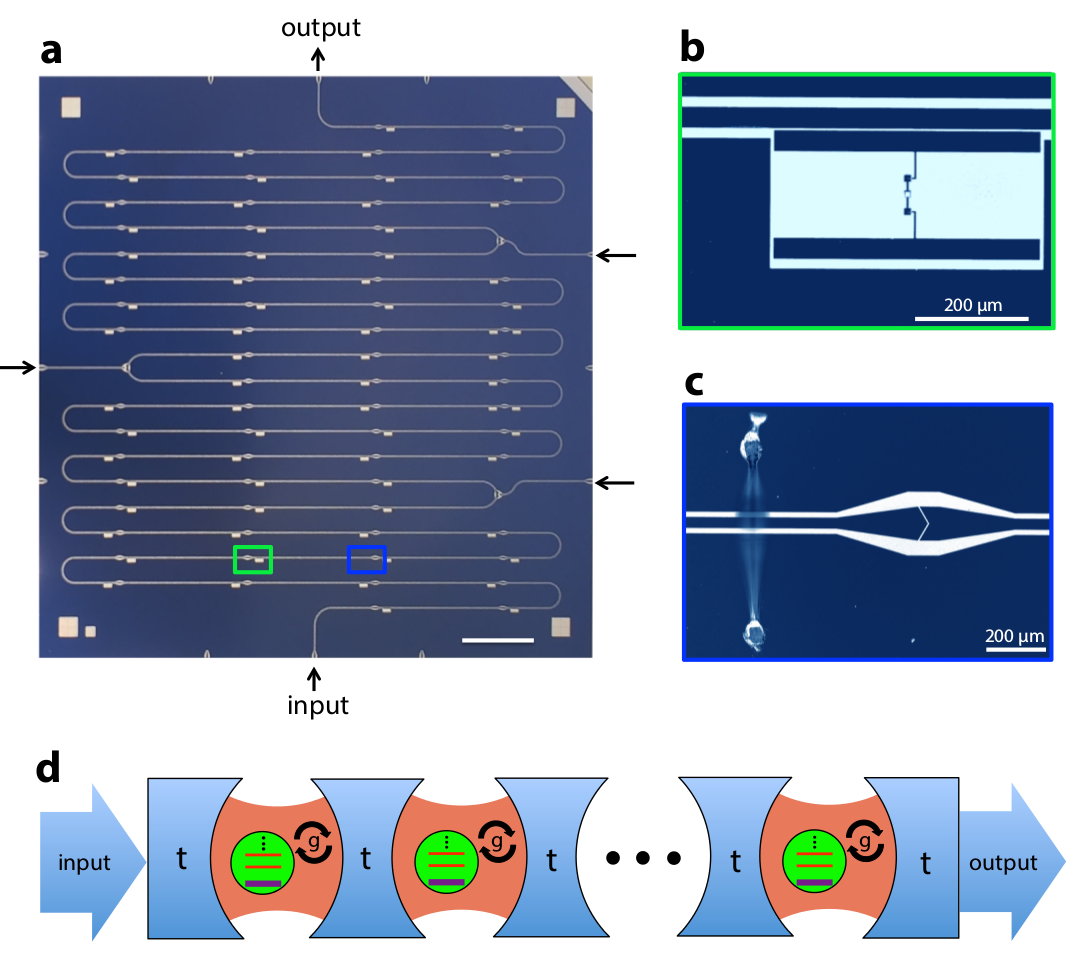}
	\end{center}
	\vspace{-0.6cm}
	\caption{\label{fig:1} 72-site circuit QED lattice. \textbf{a}, Coplanar waveguide resonators, each with a bare cavity frequency of $7.5$\,GHz, are capacitively coupled to form a linear chain on a $2.5\times2.5$\,cm$^2$ chip. Each resonator is coupled to its neighboring resonators to yield a hopping rate $t/2\pi \approx $ 144 MHz and has an average photon loss rate of $\kappa/2\pi\approx$ 1.6 MHz. At three intermediate chain sites, three-way coupling capacitors provide ports for secondary input and output lines (arrows on sides). \textbf{b}-\textbf{c}, A transmon qubit is capacitively coupled to the center pin near the edge of each resonator in the lattice, ensuring coupling to the fundamental mode of each resonator with strength $g$. The coupled resonator-qubit system forms the fundamental unit cell of the lattice. \textbf{d}, The circuit can be modeled as a linear chain of coupled oscillators, each dispersively coupled to a weakly anharmonic multi-level system.} 
\end{figure}

Our device, shown in Fig.\ 1a, consists of a linear chain of 72 lattice sites. Each site comprises a coplanar-waveguide resonator with  fundamental-mode frequency $\omega/2\pi=7.5\,$GHz, coupled to a transmon qubit\cite{Schreier2008} (Fig.\ 1b) placed at one of the resonator's voltage antinodes. 
Resonators are capacitively coupled to neighboring resonators (Fig.\ 1c), so that photons can hop between nearest-neighbor sites. Variations in transmon qubit frequencies in fabrication are a likely source of uncontrolled disorder that is difficult to compensate for in our lattice.  We therefore use an asymmetric SQUID-loop geometry allowing each qubit to be tuned over a finite frequency range via an applied magnetic flux.  Because individually tuning $72$ qubits is currently infeasible in our system,  we instead employ a global magnetic field to simultaneously tune all qubit frequencies.  Because each qubit is intentionally fabricated with a SQUID-loop of random area, this randomizes the frequency of all qubits within a band of frequencies near $8.5$\,GHz.  In this way, we can ensure that features of interest are universal to the system rather than artifacts of a particular instance of disorder (see Supplementary Information I).

\begin{figure*}[!t]
	\begin{center}
		\includegraphics[width=\textwidth]{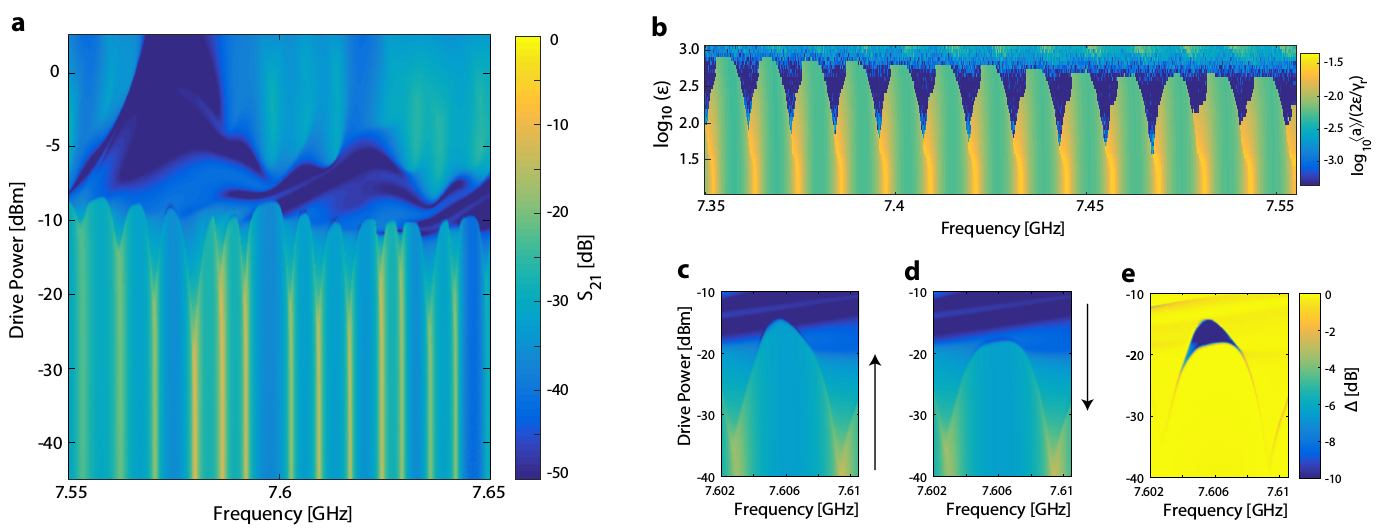}
	\end{center}
	\vspace{-0.6cm}
	\caption{\label{fig:2} Microwave transmission spectra as a function of power, exhibiting an abrupt transition to a suppressed transmission regime and a region of bistability. \textbf{a}, Dispersively shifted transmission peaks show nonlinear splitting at increased power and give rise to a region of strongly suppressed transmission without resonance peaks. Here data is acquired using constant power, frequency sweeps. \textbf{b}, Corresponding mean-field result for transmission through a 72-site lattice,  showing features qualitatively consistent with the experiment. \textbf{c}, Zoom into one lobe, showing the sharp transition to a state of suppressed transmission as the drive power is swept from low to high power using constant-frequency, linear power sweep over a $31.95$\,ms period. \textbf{d}, Same region as in \textbf{c}, but now sweeping power in the opposite direction (high to low) over the same time period as \textbf{c}. The transition now occurs along a down-shifted curve. \textbf{e}, Subtraction of the data shown in \textbf{c} and \textbf{d} uncovers the large region of hysteresis.} 
\end{figure*}

To experimentally study the nonequilibrium behavior of the device, we monitor the homodyne transmission across the lattice while varying the drive frequency and scanning the drive power over more than five orders of magnitude (Fig.\ \ref{fig:2}a). At low drive powers, we find the expected discrete transmission peaks associated with the interaction-shifted eigenmode frequencies of the resonator lattice.
As we vary the mean photon number in the system by increasing the strength of the drive, we observe that a sudden change in system behavior occurs: transmission peaks split and then, at around $-10$\,dB of drive power, abruptly give way to a region of strongly suppressed transmission. In this high-power region, peak-like features are completely absent. 

The transition between the low- and high-power phases can be more thoroughly explored by measuring the transmission at a single drive frequency while sweeping the drive power either from low to high (\ref{fig:2}c) or from high to low (Fig.\ \ref{fig:2}d).  Doing so reveals a significant region exhibiting hysteresis, which is located at the top of the low-power lobes where the transition to the high-power phase occurs. Subtracting the transmission signals for the two different sweep directions clearly marks the hysteretic regime, as shown in Fig.\ \ref{fig:2}e.

To gain insight into this behavior, we model the system as a one-dimensional chain of identical circuit QED elements, as illustrated in Fig.\ \ref{fig:1}d. The corresponding Hamiltonian 
\begin{equation}
H = \sum_j (H^\text{r}_j + H^\text{q}_j + H^\text{rq}_j) + \sum_{\langle j,j'\rangle}H^\text{hop}_{j,j'} + H^\text{d},
\end{equation}
includes terms for the resonator, qubit, and the resonator-qubit coupling on each site $j$, hopping of photons between nearest-neighbor resonators, and a coherent drive (acting on site 1 only). Each resonator contributes a single harmonic mode, $H^\text{r}_j = \omega\, a^\dagger_j a_j$, where $a^\dagger_j$ and $a_j$ are the creation and annihilation operators for photons on site $j$. The low-lying transmon levels are described as an anharmonic oscillator
$
H^\text{q}_j = \mathsf{P}_N(\Omega\, b^\dag_j b_j + \tfrac{1}{2}U\,b^\dag_j b^\dag_j b_j b_j)\mathsf{P}_N
$ 
with negative Hubbard/Kerr interaction $U=-E_C$, and projectors $\mathsf{P}_N$ that truncate the Hilbert space to levels $N\alt \sqrt{E_J/2E_C}$ within the transmon's cosine well. (Interestingly, the sign of $U$ only affects the system dynamics, but not the steady state given that the qubit and drive frequencies are tuned accordingly -- see Supplementary Information II.A.) The operators $b^\dag_j$ and $b$ create and annihilate qubit excitations, $E_C$ is the single electron charging energy and $E_J$ the effective Josephson energy of the transmon qubit. Note that we neglect disorder effects within this model. Qubit-resonator coupling and photon hopping take the simple forms $H^\text{rq}_j = g(a_j b^\dagger_j+\text{h.c.})$ and $H^\text{hop}_{j,j'} = t(a_j a_{j'}^\dag + \text{h.c.})$. Within rotating-wave approximation, the microwave drive acting on site 1 is given by $H^\text{d} = \epsilon(t)a_1 e^{i\omega_d t} + \text{h.c.}$ In our model, we account for qubit relaxation and intrinsic photon loss (at rates $\Gamma$ and $\kappa$, respectively) by employing the standard Lindblad master equation formalism for the reduced density matrix,
\begin{equation}
\dot{\rho} = -i[H,\rho] + \kappa\sum_j \mathbb{D}[a_j]\rho + \Gamma\sum_j \mathbb{D}[b_j]\rho
\end{equation} 
where $\mathbb{D}[L]\rho=L\rho L^\dag - \frac{1}{2}\{L^\dag L,\rho\}$ is the usual action of the Lindblad damping operator.
  
The experimentally observed transition is remarkably well captured by simple, quasi-classical mean-field theory that decouples the sites, but allows for mean-field parameters to differ from site to site. Allowing for site-dependent parameters is particularly relevant for our case in which the drive only acts on one end of the resonator chain, rather than on every site.) Within the quasi-classical treatment\cite{Naether2015}, the quadrature amplitudes $\alpha_j = \langle a_j \rangle$ and $\beta_j = \langle b_j \rangle$ play the role of mean-field parameters and obey the equations
\begin{align}\nonumber
i\dot{\alpha}_j &= (\omega - \omega_p - i\frac{\kappa}{2})\alpha_j  + g\,\beta_j +  t(\alpha_{j-1} + \alpha_{j+1})\ + \epsilon\,\delta_{j,1}\\
i\dot{\beta}_j &= (\Omega - \omega_p - i\frac{\Gamma}{2})\beta_j + U \abs{\beta_j}^2\beta_j + g\,\alpha_j. 
\label{MFT-eqs}
\end{align}
From these equations, we obtain the steady-state transmission signal $S_{21}\sim \langle a_j \rangle$ and the second-order coherence function $g^{(2)}(0)=\langle a_j^\dag a_j^\dag a_j a_j \rangle/|\langle a^\dag_j a_j \rangle|^2$, choosing $j$ as the label of the output port resonator (see Methods).  

The steady-state transmission (Fig.\ \ref{fig:2}b) reproduces all of the qualitative features of the experimental data.
We find that the transition occurs beyond the point where the dispersive approximation holds, and further observe that the mean-field solution predicts an increasing accumulation of transmon excitations. Higher transmon levels are a crucial model ingredient, as calculations based on the simpler Jaynes-Cummings lattice do not yield results consistent with experiment. Quasi-classically, the drop in transmission in the high-power phase is associated with chaotic dynamics
with parallels to results previously obtained for a driven, dissipative Bose-Hubbard chain\cite{Naether2015}.

While bistability obtained within nonequilibrium mean-field theory generally has to be considered with care, it is interesting to note that the mean-field solution reveals a region of bistability and hysteresis consistent with that detected experimentally (see Supplementary Information, Fig.\ 2). Finding multiple steady states appears at odds with Spohn's theorem\cite{Spohn1977}: the steady-state solution  of the Lindblad master equation is unique as long as Hilbert space is finite (or can be safely truncated), and minimal conditions for nature and number of relaxation channels are satisfied. Thus, dissipative phase transitions and stationary bistability can only occur in the thermodynamic limit -- when the number of lattice sites tends to infinity and/or when truncation fails due to accumulation of excitations in the strong drive limit, such as in the breakdown of photon blockade on a single Jaynes-Cummings site\cite{Carmichael2015, Fink2016}. 

Bistability and hysteresis can, however, be produced dynamically\cite{Drummond1980} as recently discussed by Casteels and coworkers in the context of a Bose-Hubbard dimer\cite{Casteels2016}. As shown in their  paper, hysteresis arises from parameter sweeps across a point where the spectral gap of the Liouvillian superoperator $\mathbb{L}$ (nearly) closes, in a manner analogous to the Kibble-Zurek mechanism\cite{Zurek2005}. Similar to the case studied by Casteels \emph{et al.}, we find that mean-field theory can capture certain qualitative aspects of the bistability and hysteresis, but necessarily fails in aspects related to sweep times and quantum fluctuations.

\begin{figure}[!t]
	\begin{center}
		\includegraphics[width=1.0\columnwidth]{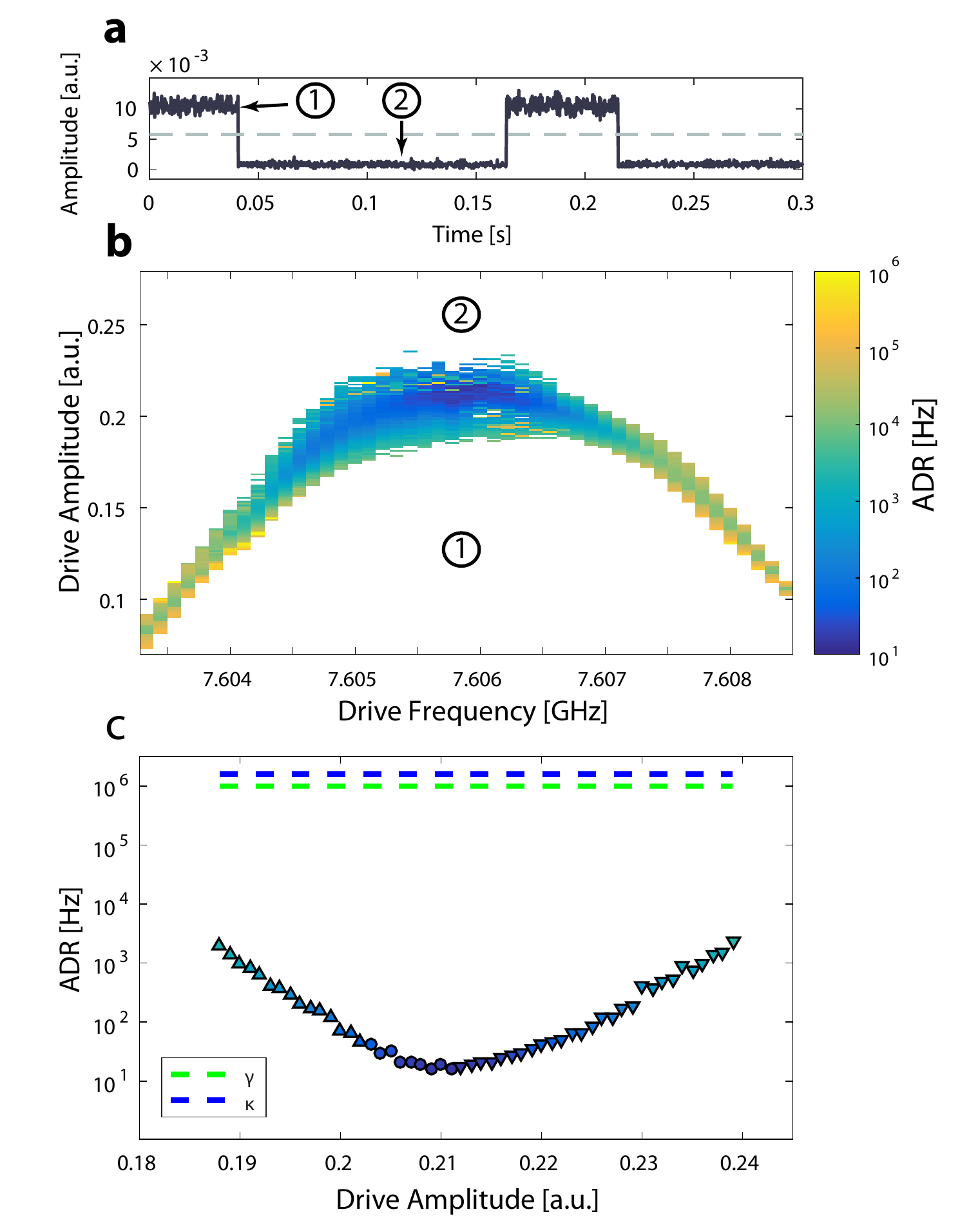}
	\end{center}
	\vspace{-0.6cm}
	\caption{\label{fig:3} Asymptotic decay rate in the transition region. \textbf{a}, Single-shot time trace of the homodyne phase in the hysteretic region for constant drive amplitude. Data show stochastic switching  between two distinct metastable states \protect\circled{1} and \protect\circled{2} on timescales vastly exceeding those intrinsic to the system.  \textbf{b}, Asymptotic decay rate obtained from the sum of the characteristic switching times, $\gamma_{1\to2}+\gamma_{2\to1}$, as a function of drive frequency and power. \textbf{c}, ADR for a drive frequency of $7.5059$ GHz is plotted as a function of power. $\kappa$ and $\gamma$ are included for reference to indicate that ADR can be as large as five orders of magnitude slower than relevant timescales of the device. When either $\gamma_{1\to2}$ or $\gamma_{2\to1}$ are slower than the duration of the measurement pulse, $\tau_m$, we cannot reliable extract a characteristic switch rate. In these cases we select the smallest extracted switching time which is larger than $1/\tau_m$. Upward (downward) pointing triangles indicate when $\gamma_{1\to2}$ ($\gamma_{2\to1}$) are less than $1/\tau_m$, circles indicate when both rates are used.}
\end{figure}

The dramatic suppression of transmission and loss of all resonance peaks beyond a certain drive power threshold are indicative of a dissipative phase transition, arising from the intricate interplay of dissipation, driving, and nonlinearity of the system.
The crucial quantity for such a transition is the gap in the spectrum of the Lindblad superoperator $\mathbb{L}$. If the real part of one of its eigenvalues approaches zero, then deviations of the steady state along the ``direction'' of the corresponding $\mathbb{L}$-eigenstate become increasingly long-lived and ultimately allow for a new steady state to emerge. The negative real part of the eigenvalue $\lambda$ closest to zero, $-\Rre \lambda$, is known as the asymptotic decay rate (ADR)\cite{Kessler2012}. An approximation for the ADR can be extracted by single-shot measurements of the dynamics in the bistable region as follows.

We apply a drive with constant frequency and amplitude, and record single-shot time traces of the homodyne amplitude and phase. Our measurements show that the system undergoes switching between two metastable states on timescales large compared to system-intrinsic timescales (Fig.\ \ref{fig:3}a).  The state of the system at each point along a single-shot trajectory is classified as either $\rho_1$ or $\rho_2$, and characteristic dwell times are extracted.  The statistics acquired from many single-shot trajectories allow us to extract average rates $\gamma_{1\to2}$, $\gamma_{2\to1}$ for the switching between the two metastable states $\rho_1$ and $\rho_2$  observed at low power and high power, respectively (labeled as \circled{1} and \circled{2} in the figure).

The extracted switching rates allow us to estimate the asymptotic decay rate by adopting a simplified rate-equation model\cite{Wilson2016} describing the probabilities $p_1$ and $p_2$ for the system to be in metastable state $\rho_1$ or $\rho_2$ (see Methods section for details):
\begin{equation}\label{equ:empirical}
\frac{d}{dt} \begin{pmatrix}
p_1 \\
p_2
\end{pmatrix} = \begin{pmatrix}
-\gamma_{1\to2} & \gamma_{2\to1} \\
\gamma_{1\to2} & -\gamma_{2\to1}
\end{pmatrix}
\begin{pmatrix}
p_1 \\
p_2
\end{pmatrix}. 
\end{equation}
Diagonalization of this system yields the stationary and purely decaying eigenmodes $\rho_s = (\gamma_{2\to1}\rho_1 + \gamma_{1\to2}\rho_2)/\gamma_\Sigma$ and $\rho_\text{ADR} = \gamma_{2\to1}\rho_1 - \gamma_{1\to2}\rho_2$ with corresponding eigenvalues zero and $\lambda_\text{ADR} = -\gamma_\Sigma = -(\gamma_{1\to2}+ \gamma_{2\to1})$. Hence, this simplified model predicts an asymptotic decay rate of $-\Rre \lambda_\text{ADR} = \gamma_\Sigma$.

Remarkably, the asymptotic decay rate, shown in Fig.\ \ref{fig:3}b-c, reaches a minimum value as low as $\sim10$\,Hz, which is five orders of magnitudes lower than the rates set by photon decay and transmon relaxation in our system. This vast timescale discrepancy delivers strong evidence for the onset of a dissipative phase transition. Similar to the situation of equilibrium phase transitions, it is only in the thermodynamic limit that the the spectral gap can fully close and turn the crossover between two steady-state phases into a phase transition in the strict sense\cite{Macieszczak2016}.

\begin{figure}[!t]
\begin{center}
\includegraphics[width=0.5\textwidth]{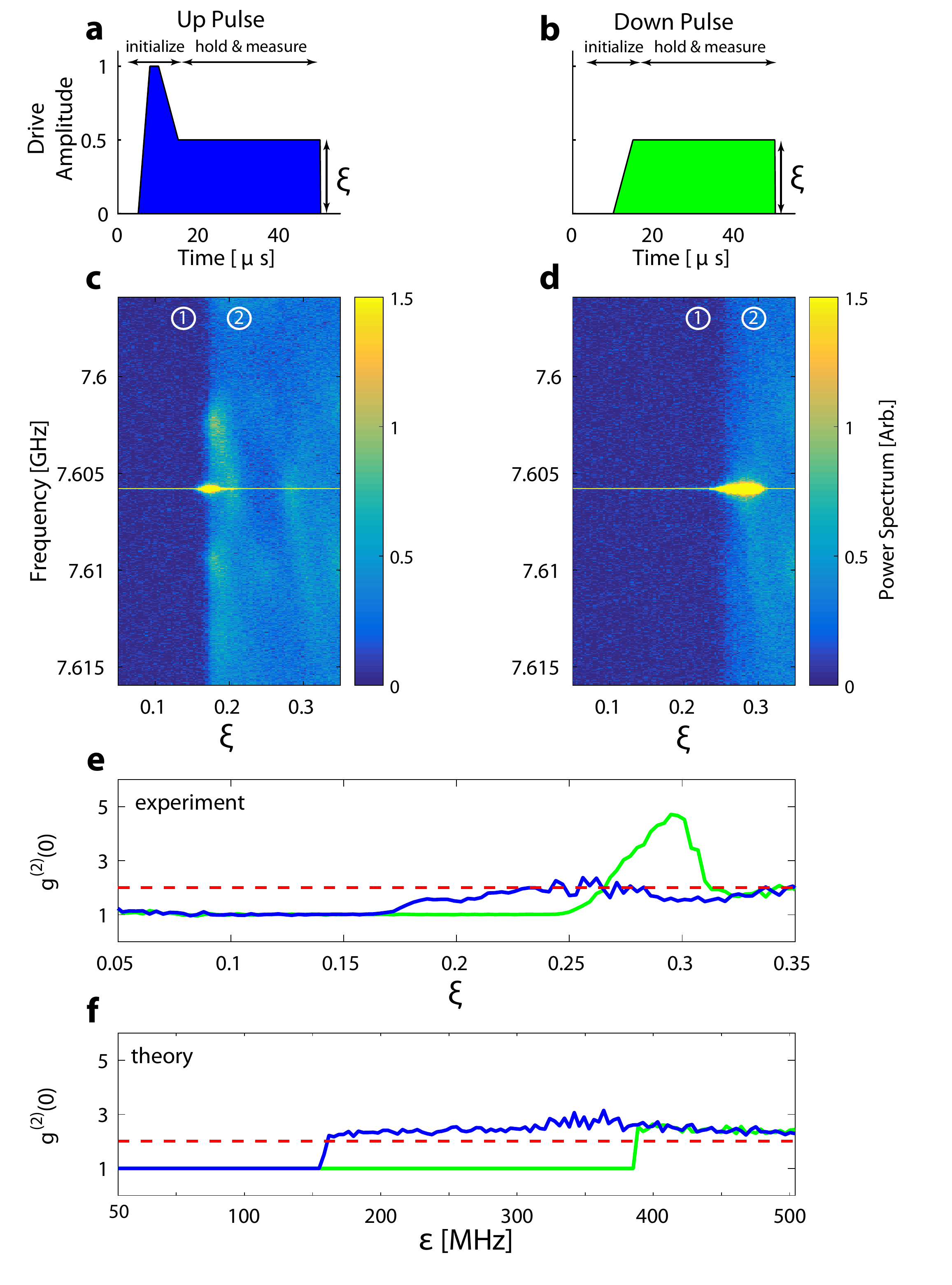}
\end{center}
\vspace{-0.6cm}
\caption{\label{fig:4} State characterization. To probe properties of the states within the region of bistability, two pulse sequences are used to initialize the system: \textbf{a},  an `up pulse'  for initialization in the high-power state, and \textbf{b}, a `down pulse' for initialization in the low-power state. Due to the long timescales, the power spectra shown in \textbf{c}, \textbf{d} and the second-order correlation function $g^{(2)}(0)$ in \textbf{e} can be obtained for each state independently, enabling state characterization within the region of bistability. \textbf{f}, Mean-field result for the second-order correlation function for comparison (note: $x$ axes cannot be compared directly).} 
\end{figure}

We gather additional evidence for the approach to a dissipative phase transition by measuring fluorescence power spectra and second-order coherence functions in our system. To this end, two different driving pulse shapes, Fig.\ \ref{fig:4}a-b, are used to access the distinct states of the system. Within the region of bistability, we can perform state initialization either in the low-power phase \circled{1} or the high-power phase \circled{2} by approaching the final drive amplitude $\xi$ either from a lower or a higher drive amplitude. After this initialization period, the two pulses maintain the constant drive amplitude $\xi$, during which time, the transmitted signal is detected using heterodyne detection with a $32$\,MHz intermediate frequency. The power spectrum is then obtained by performing a Fourier transform on the heterodyne signal. The second-order correlation function $g^{(2)}(0)$ is measured using techniques outlined in Ref.\ \cite{Eichler2012}. Figure \ref{fig:4} indicates that the low-power state can be characterized by a single, coherent drive tone ($g^{(2)}=1$) and that the high-power state can be characterized by broadband and multimode (see supplement) emission and bunching ($g^{(2)}\approx 2$). In addition, the onset of the high-power state has a stark linewidth broadening of the drive tone and has a region of strong bunching $g^{(2)}\approx 5$ for the down pulse as the system transitions to the high-power state. Experimental measurements of $g^{(2)}$ in Fig.\ \ref{fig:4}e agree well with theory results shown Fig.\ \ref{fig:4}f, barring the strong bunching observed with the `down pulse' (Fig.\ \ref{fig:4}b) at the high $\xi$ side of the bistability region.

Based on our modeling, the experiment involves both large numbers of photons and excitations of higher transmon levels, and hence may indeed approach the thermodynamic limit necessary for the observation of a dissipative phase transition.  This work demonstrates the potential for circuit QED lattices as a controllable platform that can guide a deeper theoretical and experimental understanding of nonequilibrium condensed matter physics.

\subsection*{\sffamily METHODS}
\noindent\textbf{Experimental methods.}
The cavities of the circuit QED lattice were etched using standard optical lithography and plasma etching techniques from a $200$\,nm thick Nb film on a $25\times25$\,cm$^2$ sapphire substrate. Transmon qubits were designed to have Josephson junctions with dimensions $200\times180\,$nm$^2$ and $450\times450$\,nm$^2$ and were fabricated according to the ``Manhattan" technique outlined by Potts \emph{et al.}\cite{Potts2001}, using electron beam lithography and aluminum evaporation.  Similar transmon qubits have coherence times $T_1 = 1\,\mu$s and coupling constant of $g/2\pi = 265\,$MHz.
Measurements were performed at a temperature of $7.5\,$mK in a dilution refrigerator, and inside a superconducting solenoid magnet controlled by a room-temperature DC voltage source. Transmission measurements are performed using a network analyzer, switching-rate measurements using standard homodyne detection techniques. All power-spectrum measurements were done by taking the Fourier transform of a heterodyne signal, and $g^{(2)}$ measurements were implemented using the homodyne techniques described by Eichler \emph{et al.}\cite{Eichler2012} (see Supplementary Information for further details).\\

\noindent\textbf{Numerical solution of the mean-field equations.}
We solve for the stationary state of the mean-field equations \eqref{MFT-eqs} by time evolution and extracting the long-time limit, since root-finding methods are difficult to handle for the large system of nonlinear equations\cite{Naether2015}. In the high-power phase the dynamics is chaotic, so that additional time averaging in the long-time limit is required. For instance, the second-order coherence function is obtained by evaluating
$
g^{(2)}(0) = \langle\!\langle |\alpha(t)|^2 \rangle\!\rangle_t / |\langle\!\langle |\alpha(t)| \rangle\!\rangle_t|^2,
$
where the time average $\langle\!\langle\cdot\rangle\!\rangle_t$ is carried out over a time interval that excludes any initial transient behavior.\\

\noindent\textbf{Model underlying the ADR estimate.}
First, consider stochastic switching between two pure states $\ket{1}$ and $\ket{2}$. The simplest description is based on a two-level Hamiltonian
$
H = 
E_{21} \left| 2 \right> \left< 2 \right| 
$
where $E_{21}$ is the energy difference between the two states, and the master equation
\begin{equation}
\label{eq:master_eqn}
\dot{\rho} = -i \left[H, \rho \right] + \gamma_{1\to2} \mathbb{D} \left[  \left| 2 \right> \left< 1 \right| \right]\rho + \gamma_{2\to1} \mathbb{D} \left[  \left| 1 \right> \left< 2 \right| \right]\rho,
\end{equation} 
with $\mathbb{D}[L]$ denoting the usual Lindblad damping superoperator for jump operator $L$. The resulting $4\times4$ Liouvillian $\mathbb{L}$ is block-diagonal, where one of the two blocks fully captures the dynamics of density matrices of the form $\rho(t)=p_1(t)|1\rangle\langle1| + p_2(t)|2\rangle\langle2|$, where the probabilities $p_{1,2}$ obey the rate equation \eqref{equ:empirical}.
This model can be extended and made more realistic by considering subsets of pure states that make up the two metastable states $\rho_1$ and $\rho_2$, which are likely to be mixed states rather than pure states.

\subsection*{Acknowledgments}

The authors thank Iacopo Carusotto and Cristiano Ciuti for helpful discussions. This work was supported by  the Army Research Office through grant W911NF-15-1-0397 and the National
Science Foundation through Grants No. DMR-0953475 and No. PHY-1055993.  NS was supported by an NDSEG fellowship.

\vfill

\newpage
\begin{center}\Large{Supplemental Information}
\end{center}

\section{Device Parameters}\label{sSec:deviceparameters}
The device consists of 72-coplanar waveguide cavities coupled by  capacitors which are formed using gaps in the center pin of the resonator. Each tunneling capacitor is designed to have a capacitance of $20.7$ pF, resulting in a tunneling matrix element $t/2\pi = 144$ MHz. The three-way couplers are integrated in the lattice to maintain the cavity-cavity hopping rate while introducing a weak coupling to an input/output transmission line to weakly probe the internal behavior of the lattice. 

\begin{figure*}[!t]
\begin{center}
\includegraphics[width=0.9\textwidth]{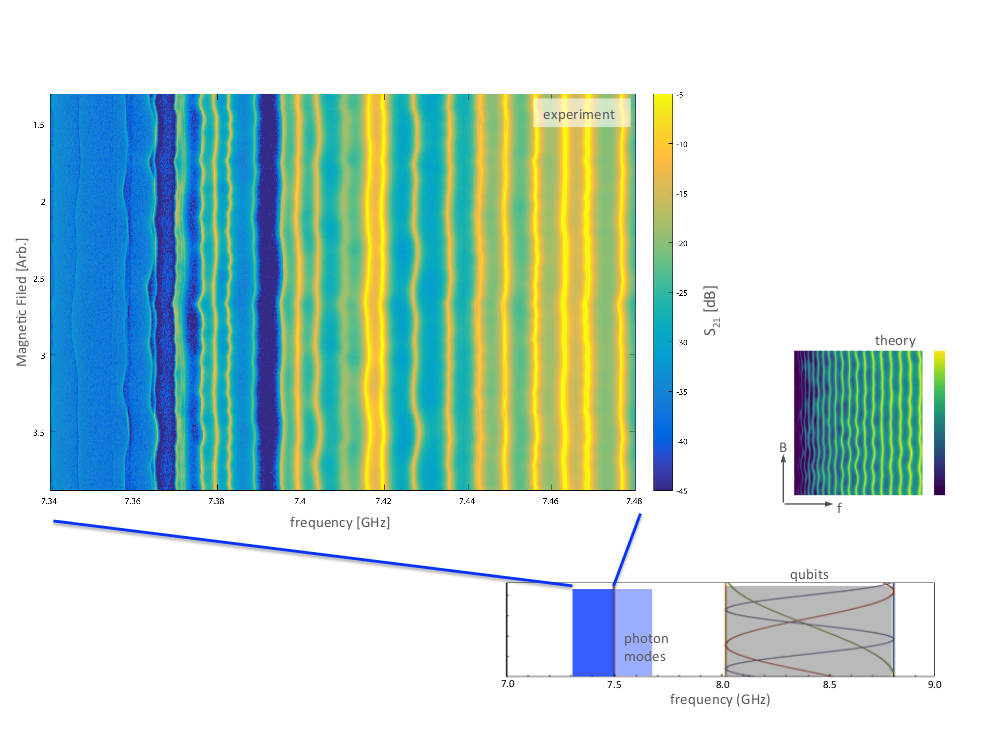}
\end{center}
\vspace{-0.6cm}
\caption{\label{fig:s1} Transmission as a function of external magnetic field. By changing the strength of the external magnetic field, individual qubit frequencies can be continuously changed. The non-periodic behaviour of the transmission peaks is due to the random areas of the transmons in the cQED lattice.} 
\end{figure*}

One qubit is coupled to each cavity.  From finite-element simulations, we predict a charging energy $E_c/h=180$ MHz. Similar qubits measured in other devices have measured coupling rates $g/2\pi = 265$ MHz.  Each transmon SQUID has a fixed width of 9 $\mu$m but the heights are chosen to be a random number between 8 $\mu$m and 20 $\mu$m, ensuring that a global magnetic field  can continuously change the qubit frequencies but will never return to the exact same qubit frequency distribution.  Together with previous measurements of $E_J$ and simulated values for $E_c$, we expect qubit frequencies to be between $8$ and $8.8$ GHz. This is confirmed in Fig. \ref{fig:s1} which shows a measurement of transmission as a function of external magnetic field. Each mode exhibits frequency shifts of different flux periodicity, indicating the existence of numerous qubits. Based on optical microscope inspection of the sample, it is expected that after the experiment was complete, a minimum 60\% of our qubits are fully functional, with another 20\% for which the smaller Josephson junction of the SQUID did not make a physical connection; it is unknown if this damage occured due to handling the device after the experiment was complete.

\newpage
\section{Theory}\label{sSec:theory}
	
	 \subsection{Mapping of positive -- negative $U$}
	 The sign of the Hubbard interaction $U$ has a significant impact on the system's energy spectrum. However, if we are merely interested in the steady-state behavior, the sign of $U$ is relatively less important as long as the system remains in the transmon's cosine well. We begin our discussion first with the quasi-classical treatment. For the steady state, the time derivative of any expectation value vanishes. Hence, the quasi-classical dynamical equations reduce to
	 \begin{align}\nonumber
	 0 &= (\omega - \omega_p - i\frac{\gamma}{2})\alpha_j  + g\,\beta_j +  t(\alpha_{j-1} + \alpha_{j+1})\ + \epsilon\,\delta_{j,1},\\
	 0 &= (\omega + \Delta - \omega_p - i\frac{\Gamma}{2})\beta_j + U \abs{\beta_j}^2\beta_j + g\,\alpha_j,
	 \end{align}
	 with the qubit-resonator detuning $\Delta = \Omega - \omega$. Now, we consider a mapping such that
	 \begin{equation}\nonumber
	 \left(\omega_p , \Delta, g, t, \alpha_j, \beta_j \right) \rightarrow
	 \left(2\omega - \omega_p, - \Delta, -g, -t, \alpha_j^*, \beta_j^* \right).
	 \end{equation}
	 This mapping physically corresponds to tuning both the drive and qubit frequencies in a specific way. (For our open chain, the sign flip of $g$ and $t$ is allowed by a local gauge transformation.) By taking the complex conjugate on both sides of the dynamical equations together with the mapping, we find:
	 \begin{align}\nonumber
	 0 &= (\omega - \omega_p - i\frac{\gamma}{2})\alpha_j  + g\,\beta_j +  t(\alpha_{j-1} + \alpha_{j+1})\ + \epsilon\,\delta_{j,1},\\
	 0 &= (\omega + \Delta - \omega_p - i\frac{\Gamma}{2})\beta_j + (-U) \abs{\beta_j}^2\beta_j + g\,\alpha_j.
	 \end{align}
	 The mapping thus effectively flips the sign of $U$. In other words, if both the drive and qubit frequencies are tunable, the sign of $U$ is not a concern and the truncation to low-lying transmon levels is justified.
	 
	 The above discussion can be easily generalized to the Lindblad master equation. In that case, instead of mapping $\alpha_j$ and $\beta_j$ to their complex conjugates, we map the steady-state density matrix $\rho_{s}$ to its complex conjugate $\rho_{s}^*$. Note that any steady-state expectation value can be calculated using $\rho_{s}^*$ through the equation
	 \begin{equation}
	 \langle O \rangle = \mathrm{tr} \left( O \rho_{s} \right) =\left[ \mathrm{tr} \left( O^* \rho_{s}^* \right) \right]^*.
	 \end{equation}
	 
\subsection{Simulation of hysteresis}
\begin{figure}[!t]
\begin{center}
\includegraphics[width=0.5\textwidth]{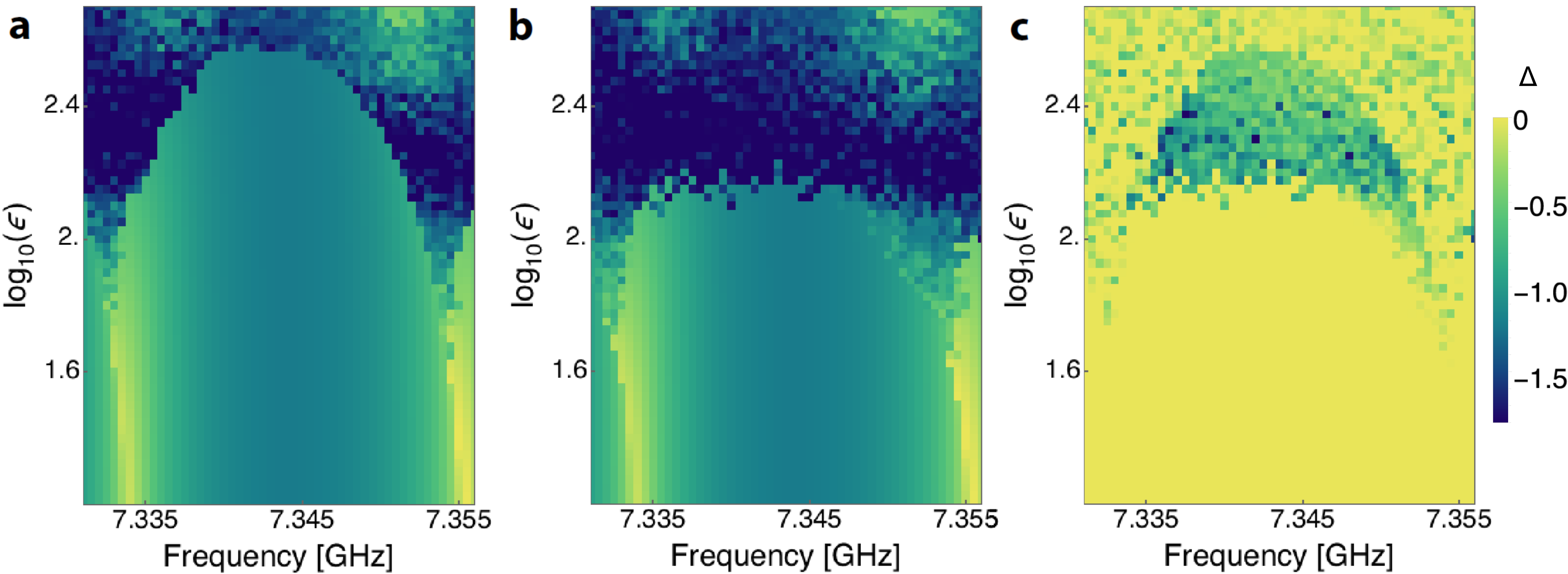}
\end{center}
\vspace{-0.6cm}
\caption{\label{fig:s_sim_hysteresis} Simulation of the region of hysteresis. Transmission of a 20-site chain is simulated with parameters motivated by the experiment, and initial state being the vacuum state in \textbf{a} and a highly excited state in \textbf{b}. \textbf{c}, Subtraction of the simulated transmission shown in \textbf{a} and \textbf{b} uncovers the region of hysteresis.} 
\end{figure}

Hysteresis appears in the simulation if the quasi-classical dynamical equations have more than one attractor. Hence, we can numerically investigate the region of hysteresis by picking different initial states and checking whether the system evolves to the same or different attractors in the long-time limit. We illustrate this by simulating the transmission for a 20-site chain in Figure \ref{fig:s_sim_hysteresis}. (The region of hysteresis is qualitative the same for both 20-site and 72-site chains.) Subtraction of the simulated transmission with two different initial states shows a region with more than one attractor. The shape of this hysteresis region is consistent with the experimental data. Readers may notice that the result becomes 'noisy' in the high power regime. This is the result of the chaotic dynamics which makes the time average strongly sensitive to the initial state and the averaging window. However, the hysteresis is reassuring by direct inspection of the system time evolution. In the region of hysteresis, one initial state gives stationary behavior (stable fixed point) in the long time limit while the other gives chaotic dynamics (strange attractor).

\newpage
\section{Switching Rate Extraction}\label{sSec:jumpRateExtraction}
\begin{figure}[!t]
\begin{center}
\includegraphics[width=0.5\textwidth]{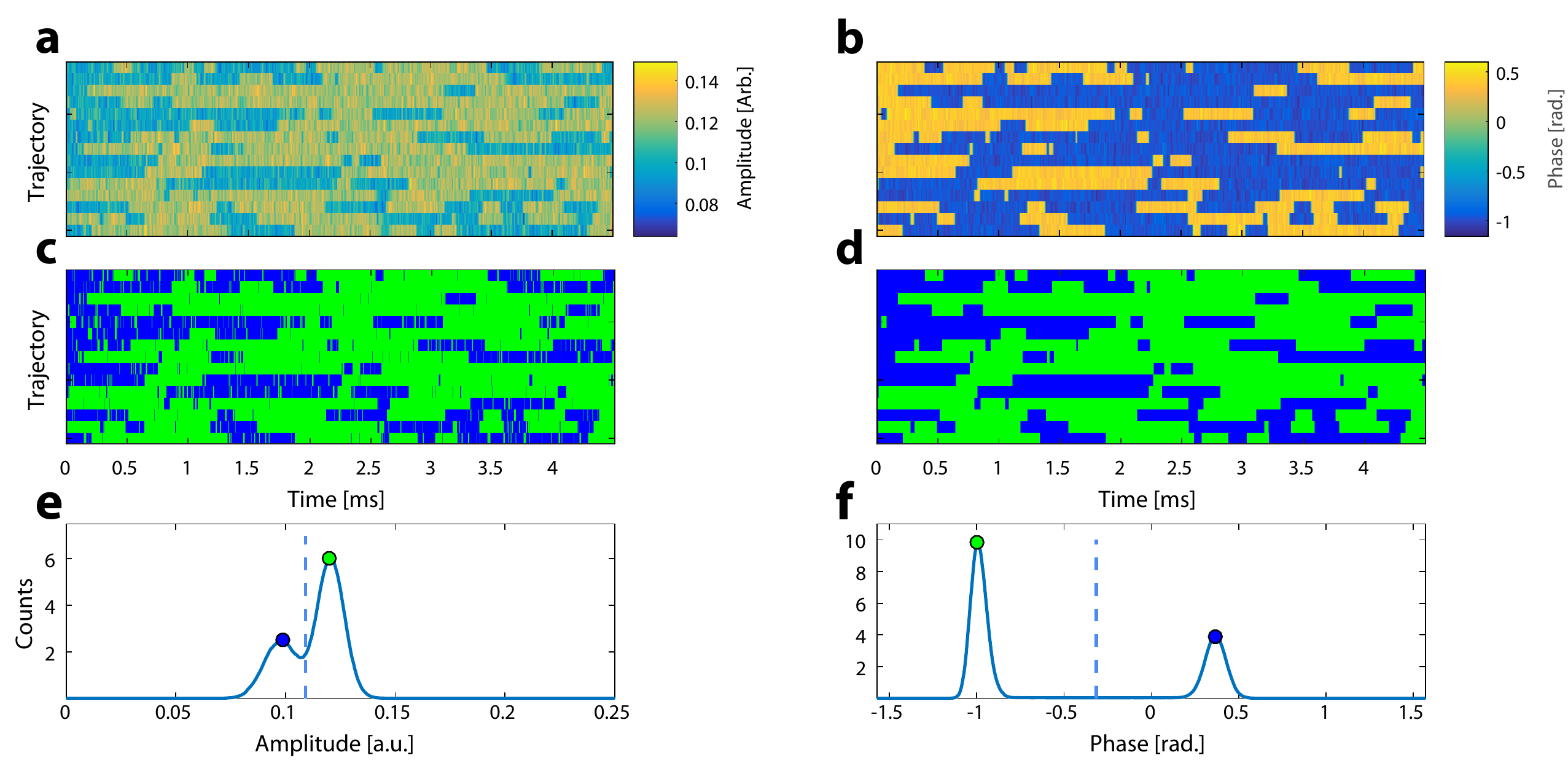}
\end{center}
\vspace{-0.6cm}
\caption{\label{fig:s2} Thresholding Procedure. Single-shot trajectories of the homodyne amplitude [(a),(c),(e)] and phase [(b),(d),(f)] demonstrate switching between two distinct steady states of the system. From the raw data acquired in (a) and (b), a histogram is compiled and, if the resulting distribution is bimodal, the mean of the peak locations is used as a discriminating threshold.  Each point along a trajectory is categorized based on the threshold, resulting in (c) and (d). Ultimately, the state lifetimes are determined using the threshold which has the fewest histogram counts at the threshold (phase in this case).} 
\end{figure}

In order to extract switching rates necessary for the calculation of the asymptotic decay rate (ADR), it is necessary to gather statistics on the lifetimes of each state in the region of bistability. To do this, a pulse shown in Fig. \ref{fig:4}, lasting 0.3 s is used to modulate a continuous-wave (CW) microwave source. After the initialization phase of the pulse, the homodyne amplitude and phase are extracted from the output signal by first passing the raw signal through a series of amplifiers before being mixed down to using a local oscillator (LO) at the drive frequency using an IQ mixer. The I and Q outputs of the IQ mixer are then sent through a $1.9$ MHz low-pass filter (LPF) before being amplified, digitized, and sent to the measurement computer (see Fig. \ref{fig:s2}). The amplitude and phase of the homodyne signal are then extracted by taking

\begin{align}\label{equ:homodyne}
A&=\sqrt{I^2 + Q^2}\\
\theta&=\tan^{-1}(I/Q)
\end{align}

The $1.9$ MHz filter was chosen to reduce noise in the single-shot signal in an effort to reduce the number of false-counts in the steady state transitions. This filter frequency is well above observed transition rates.  The digitizer sampling rate was chosen to be 50 MS/s and the data was down-sampled to include only every tenth data point to avoid memory constraints. After all processing, each trajectory consists of $1.5 \times 10^{6}$ data points.

\begin{figure}[!t]
\begin{center}
\includegraphics[width=0.5\textwidth]{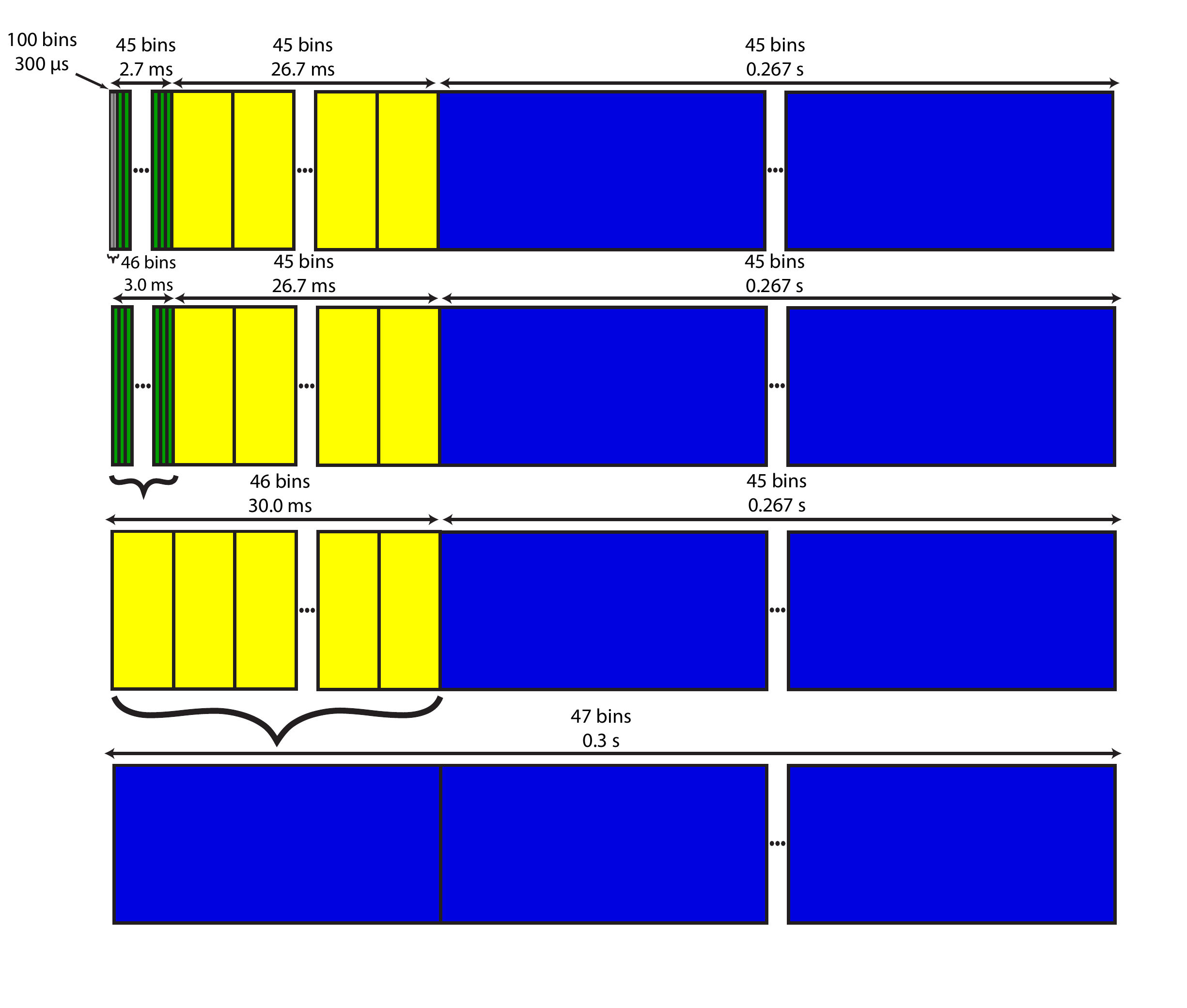}
\end{center}
\vspace{-0.6cm}
\caption{\label{fig:s3} Switching time histogram bins. When transition times are short, the short time bins provide a precise determination of short characteristic switching times. When the switching times are long, however, the shorter time interval bins can be summed to form larger bins, creating well-populated bins of larger time intervals, enabling a precise determination of longer characteristic switching times.} 
\end{figure}

Figure \ref{fig:s2} outlines the thresholding algorithm used to determine switching rates $\gamma_{1\to2}$ and $\gamma_{2\to1}$. For each drive amplitude and frequency, seven 0.3 s trajectories are acquired and the data is placed in histograms for amplitude and phase. It is then determined whether the resultant distributions are gaussian, meaning there is no bistability, or bimodal, meaning that the there is bistability. In the case of a bimodal distribution, data are classified as either being in the high or low-power state with a threshold given by the mean of the peak locations. The measured quantity (amplitude or phase) which has the fewest histogram counts at the threshold is then used as the measured quantity for the state lifetime determination. New data is then acquired and categorized according to the threshold. Once the new data has been categorized, state lifetimes are extracted. The lifetimes are then binned in a nonuniform histogram bins shown in Fig. \ref{fig:s3}.

Depending on the population of the histogram bins, the bins can be summed to create histograms with larger time intervals. This scheme ensures that histograms are sufficiently populated for situations involving short or long lifetimes. After the binning procedure has been performed, the resultant distribution is fit to an exponential to extract a characteristic switching time, $\tau_{c}$. From this, the switching rate for the state of interest is $\gamma=1/\tau_{c}$.   

The physical measurement setup for the jump rate extraction is shown in Fig. \ref{fig:fridge}.

\section{Emission Properties}
The transmission shown in Figure 2 of the main text exhibits an abrupt change as the system crosses into the high power regime. 
As illustrated in Figure 4 of the main text, the high power state can be characterized by broadband emission around the drive tone. When a power spectrum measurement is performed (in this case using a spectrum analyzer) observing the entire range of frequencies with low-power transmission peaks, the system emits at nearly all the low-power transmission peaks of the system. This can be understood qualitatively by considering the system Hamiltonian in the eigenmode basis of the cavity chain.

\begin{figure}[!t]
\begin{center}
\includegraphics[width=0.5\textwidth]{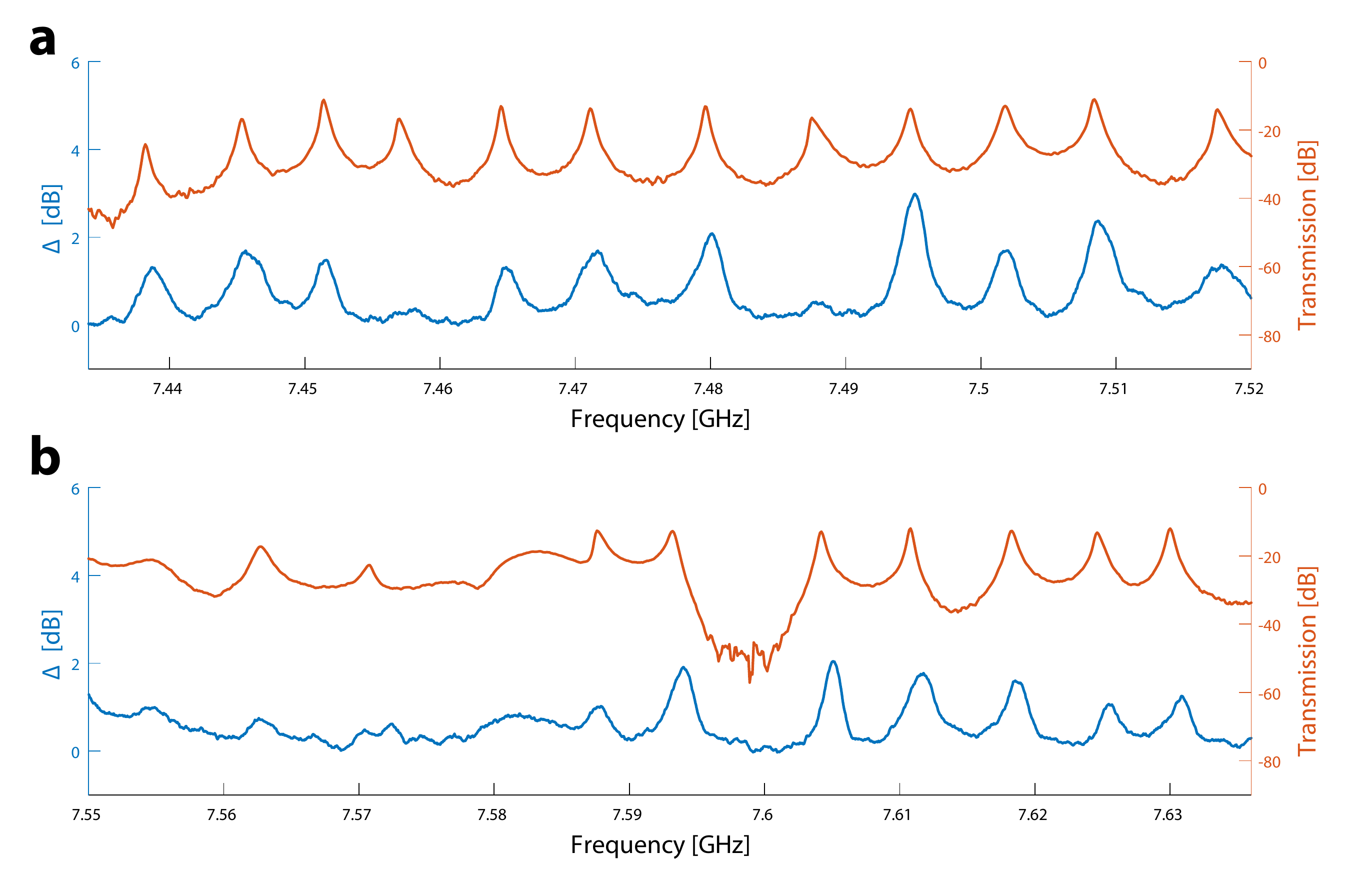}
\end{center}
\vspace{-0.6cm}
\caption{\label{fig:s6} Multimode Emission. By driving a single mode of the system, in this case at 7.535 GHz, emission is observed both below (a) and
above (b) the drive tone. Emission peaks (shown in red) qualitatively match low-power transmission peaks (shown in blue). Emission is plotted in units of dB above the background.} 
\end{figure}

\begin{figure*}[!t]
\begin{center}
\includegraphics[width=\textwidth]{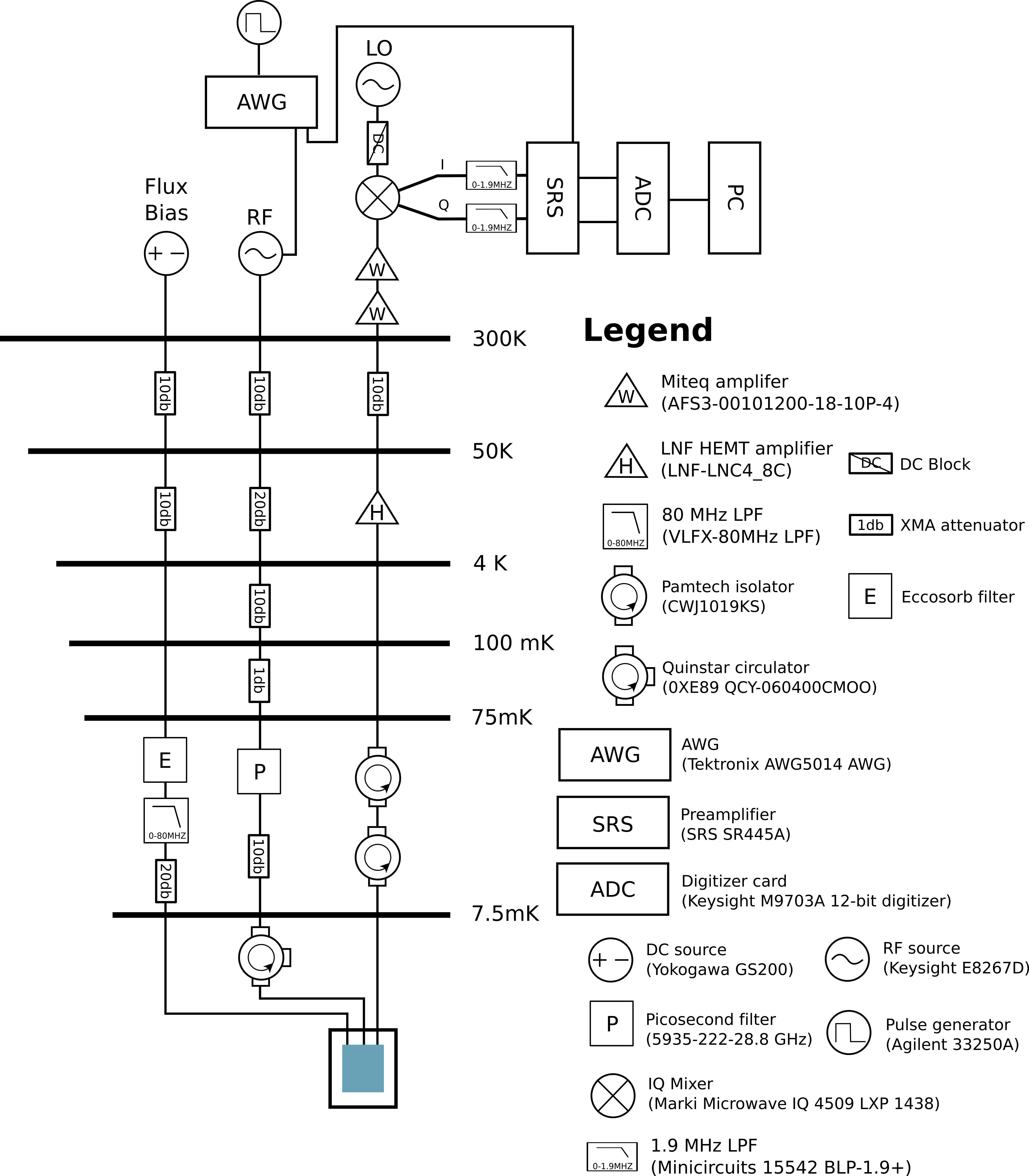}
\end{center}
\vspace{-0.6cm}
\caption{\label{fig:fridge} Measurement schematic for extraction of switching times. Measurements are performed inside a superconducting magnet at the base of a dilution refrigerator. Transmission measurements shown in Fig. \ref{fig:2}
 are performed using the same experimental setup but using a network analyzer (Keysight PNA-X N5241A) to source the RF signal and the returning signal after the two Miteq amplifiers is sent directly back into the network analyzer.} 
\end{figure*}

We can rewrite the Hamiltonian in the cavity eigenmode basis by the substitution: $a_{j}=\sum_{\mu=1}^{N} W_{j \mu} \tilde{a}_{\mu}$, where $W_{j\mu}$ is the weight of the $\mu$th eigenmode at site $j$. This gives
\begin{align}\label{equ:dicke2}
H = &
\sum_{\mu=1}^{N}\tilde{\omega}_\mu \tilde{a}_{\mu}^{\dagger} \tilde{a}_\mu +
\sum_{j=1}^{N} \left( 
\Omega b_j^{\dagger} b_j  
+ \frac{U}{2} b_j^{\dagger} b_j^{\dagger} b_j b_j
\right)
+ \\
& g \sum_{j,\mu=1}^{N} \left(
W_{j \mu}^{*} \tilde{a}_{\mu}^{\dagger} b_j + \text{h.c.}
\right)
 +\epsilon\sum_{\mu=1}^{N} \left( W_{1\mu}^* \tilde{a}_{\mu}^{\dagger} + \text{h.c.}\right), \nonumber
\end{align}
where $\tilde{\omega}_{\mu}$ is the frequency of the $\mu$th eigenmode. This equation can be thought of as a collection of modes coupled to a bath of transmon qubits.  In this basis, each cavity eigenmode interacts with all other cavity eigenmodes through the communal qubit bath. This effective interaction explains the presence of multimode emission shown in Figure \ref{fig:s6}.

\newpage
%

\end{document}